\documentclass[11pt]{revtex4}

\usepackage{graphicx}
\usepackage{dcolumn}
\usepackage{color}
\usepackage{bm}
\usepackage{amsmath}
\usepackage{amsfonts}
\usepackage{amssymb}
\usepackage{undertilde}
\usepackage{epstopdf}

\input epsf

\begin{document}

\title{A Comparative Study of Different Entropies in Fractal Universe}

\author{Sourav Haldar\footnote {sourav.math.ju@gmail.com}}
\affiliation{Department of Mathematics, Jadavpur University, Kolkata-700032, India.}

\author{Jibitesh Dutta\footnote {jdutta29@gmail.com,jibitesh@nehu.ac.in}}
\affiliation{Mathematics Division,Department of Basic Sciences and
Social Sciences, North Eastern Hill University, Shillong 793022,
Meghalaya, India.}

\author{Subenoy Chakraborty\footnote {schakraborty@math.jdvu.ac.in}}
\affiliation{Department of Mathematics, Jadavpur University, Kolkata-700032, West Bengal, India.}

\begin{abstract}
Here we make an attempt to extend the idea of generalized Hawking
temperature and modified Bekenstein entropy at event horizon in
fractal universe.  The modified Hawking temperature and Bekenstein
entropy is  considered in the governing Friedmann equations, which
 is modified in the background of a fractal universe. The validity
 of the Generalized second law of thermodynamics\,(GSLT) and Thermodynamic
 Equilibrium\,(TE) have been examined and compared graphically in a
 fractal universe filled with perfect fluid having constant equation
 of state in three different generalized Bekenstein system.

\end{abstract}

\maketitle

\section{Introduction}

In 1970s Hawking and Bekenstein discovered thermodynamics of black holes\,(BH).
 Since then there is a general belief that there is a profound connection between
 gravity and thermodynamics \cite{Bekenstein1,Hawking1,Bekenstein2,Bardeen1,Hawking2}.
 Eventually the BH are considered to be a black body emitting thermal radiations
 known as Hawking radiation \cite{Bardeen1,Hawking2}. It was realized that laws
 of BH physics and thermodynamical laws are equivalent. Furthermore, Jacobson and Padmanabhan did a pioneer work in this
direction by establishing a relationship between first law of
thermodynamics and Einstein field equations\cite{Jacobson1,
Padmanabhan1}. Afterwards,  these results are generalized in the
cosmological background  and there have been lot of studies
dealing with universe as a thermodynamical
 system \cite{Akbar1}.

In the context of Universal thermodynamics, usually universe is
considered to be bounded by apparent horizon. However, in
accelerating universe the apparent horizon is different from the
event horizon.  The cosmological event horizon usually does not
exist in standard Big Bang Cosmology but assured to exist in
accelerating phase.
 In this context Wang \textit{et
al} \cite{Wang1,Zhou1} in 2006 investigated the laws of thermodynamics in an
 accelerating universe dominated by Dark Energy with a time dependent equation
 of state. They showed that the first and second laws of thermodynamics are satisfied
 on apparent horizon while the thermodynamical laws break down on cosmological event
 horizon. As a result, they concluded that the universe bounded by apparent horizon
 is a Bekenstein system\,(perfect thermodynamical system) and termed the universe
 bounded by event horizon as a non Bekenstein system\,(an unphysical system).
 Later it has been shown that the generalized second law of thermodynamics holds (in any gravity theory)
 with some reasonable restrictions for the Universe bounded by an event horizon under the assumption that the first law holds for Einstein gravity \cite{Mazumder1,Mazumder2}
 and in other gravity theories \cite{Mazumder1,Mazumder2,Mazumder3} and for different fluid systems \cite{Mazumder1,Mazumder2,Chakraborty1} (including dark energy
 \cite{Mazumder2,Chakraborty1}).  Furthermore, due to complicated
 nature of event horizon there are very few studies related to
 event horizon. So it is natural to investigate validity of
 thermodynamical laws on event horizon.

 The usual definition of thermodynamical parameters do not serve the purpose on event horizon as universe is non static in this case. In recent past, Chakraborty and Saha\cite{Chakraborty2, Chakraborty2a} introduced
 generalized Hawking temperature and modified Bekenstein entropy on event Horizon and they proved the validity of thermodynamic laws
 on event horizon without assuming first law in the Einstein gravity
 \cite{Chakraborty3}. As thermodynamic interpretation of gravity
 near horizon is generic feature, it is imperative to verify
 thermodynamical laws in more general space-time. Here, we would
 like to see the validity of thermodynamical laws in fractal
 universe with alternative definition of thermodynamical
 parameters.
Historically the first appearance of fractal cosmology was in
Andrei  Linde's paper \cite{Linde:1986fd}. For an overview of
fractal cosmology one can see the ref \cite{Jonathan}. Later,
  Calcagni \cite{Calcagni1,Calcagni2} made a theoretical approach for a power-counting renormalizable
field theory living in a fractal space-time and consequently fractal cosmology was
 developed.  The action in this model is Lorentz covariant and the
 space-time metric ($\mathcal{M}, \varrho$) is equipped with a Stieltjes measure
 $\varrho$. Very recently it was shown that, the Friedmann equations can be transformed
 to Clausius relation, but a treatment with non-equilibrium thermodynamics of space-time
 is needed \cite{Haldar:2015nna}. Furthermore, Sheykhi \textit{et al} examined GSLT
 in a fractal universe on apparent horizon and found that GSLT is valid for particular
 choice of fractal parameter \cite{Sheykhi4}.
 In this letter, we make an attempt to compare three different generalized Bekenstein formulation on event horizion in Fractal Universe and examine  the validity of thermodynamical laws
 with  these generalised definition of thermodynamical parameters.
  The main focus is to extend the idea of
generalized Hawking temperature and modified Bekenstein entropy in
fractal universe. A similar comparative study of alternative
thermodynamical  parameters in $f(R)$ gravity  has been performed
in recent past\cite{Dutta:2016sgj}.
 The paper is organized as follows : Section II deals with basic concepts related to
 earlier works. Brief review and basic equations of fractal cosmology are presented in
 section III, while section IV deals with thermodynamical analysis in this context. Finally,
 summary of the work and possible conclusions are presented in section V.

\section{Basic equations of universal thermodynamics}

Let us consider the homogeneous and isotropic FRW model of universe expressed by the metric
\begin{equation}\label{eq1}
ds^2= h_{ij}(x^i)dx^{i}dx^{j}+ R^{2}d\Omega ^2 _2
\end{equation}
where $h_{ij}= \mbox{diag}\left(-1 , \frac{a^2}{1-kr^{2}}\right)$ is the two-dimensional
 metric tensor, known as normal metric. Here $x^0= t , x^1= r~~~i.e.~~ i , j$ can take
 values 0 and 1, $R= ar$ being the area radius  considered in the normal 2-D space. On this
normal space another relevant scalar quantity is defined as
\begin{equation}\label{eq2}
\chi(x)= h_{ij}\,\partial _{i}R\,\partial _{j}R\\= 1- \left(H^2+ \frac{k}{a^2}\right)R^2
\end{equation}
where $k= 0 , +1 , -1$ stands for flat, closed or open model, and $H= \frac{\dot{a}}{a}$ is the Hubble parameter.

The apparent horizon is given by the vanishing of the scalar $\chi(x)$ as
\begin{equation}\label{eq3}
R_A= \frac{1}{\sqrt{H^{2}+\frac{k}{a^2}}}
\end{equation}
which becomes $\frac{1}{H}$ for a flat space ($i.e.~~k= 0$). On
the other hand,
 event horizon\,(EH) is given as $R_E= a\int _{t}^{\infty} \frac{dt}{a(t)}$ which  exists only
 in the present accelerating era.


The generalized second law of thermodynamics\,(GSLT) states that
the total entropy of an isolated
 macroscopic physical system should be a non-decreasing function and ultimately  such a system always
 evolves towards thermodynamic equilibrium\,(TE). So the following two inequalities can be used
 to verify the validity of GSLT and TE :
\begin{eqnarray}\label{eq4}
\textrm{for GSLT :}~~~\frac{dS_{T}}{dt}\geq 0\nonumber\\
\textrm{and for TE :}~~~\frac{d^{2}S_{T}}{dt^{2}}< 0
\end{eqnarray}

where $S_T= S_{h}+ S_{f}$, with $S_h$ and $S_f$ denoting respectively the horizon entropy and
 the entropy of the fluid bounded by the horizon. One can use Gibb's relation to calculate $S_f$,
\begin{equation}\label{eq5}
T_{f}dS_{f}= dE_{f}+ pdV_{h}
\end{equation}
where $V_h$ is the volume of the fluid, $E_{f}= \rho V_h$ is the total energy of the fluid and
 $T_f$ is the temperature of the fluid.

In the present context, it is assumed that the temperature $T_f$ of the cosmic fluid inside the
 horizon is same as that of the bounding horizon {\it i.e.} $T_h$ , unless there is a spontaneous
 flow of energy between the horizon and the fluid which is not consistent with FRW model. So it
 is assumed that $T_{f}\propto T_h$ or $T_{f}= bT_h$, which is widely taken as $T_{f}= T_h$ to
 avoid mathematical complexity of non-equilibrium thermodynamics.

\section{Basic Equations Of Fractal Universe}

The fractal properties of quantum gravity theories in $n$-dimensions have been explored in several
 contexts. At first, renormalizability of perturbative gravity at and near two topological dimensions
 drew much interest into $n= 2+ \epsilon$ models, with the hope to understand the $n= 4$ case better.
 Assuming that matter is minimally coupled with gravity, the total action of Einstein gravity in a
 fractal space-time is given by,
\begin{equation}\label{eq6}
S= S_{G}+ S_m
\end{equation}
where
\begin{equation}\label{eq7}
S_{G}= \frac{M^2 _P}{2}\int d^{n}x\sqrt{-g}\left(R - u\,\partial _{\mu}v\,\partial ^{\mu}v\right)
\end{equation}
is the gravitational part of the action and
\begin{equation}\label{eq8}
S_{m}= \int d^{n}x\sqrt{-g}\mathcal{L}_m
\end{equation}
is the matter part of this action. Here $g$ is the determinant of the metric $g_{\mu \nu}$,
 $M_{P}= (8\pi G)^{-1/2}$ is reduced Planck mass and $R$ is the Ricci scalar. Also $u$ and
 $v$ respectively denote the fractal parameter and the fractal function\,(plays the role of
 a weight function of the integral in eq.\,(\ref{eq8})) respectively.


 The standard measure in the action is replaced by a nontrivial measure which appears in Lebesgue
-Stieltjes integral $$d^{n}x= d \varrho (x).$$ The scaling dimension of $\varrho$ is $[\varrho]=
 -n\alpha _{f} \neq -n$, where $\alpha _{f} > 0$ is a positive parameter.\\\\
Taking the variation of the action (\ref{eq6}) with respect to the FRW metric $g_{\mu \nu}$, and assuming $8\pi G$ to be unity for convenience, for a flat FRW metric, the Friedmann equations can be obtained in a fractal universe as
\begin{equation}\label{eq9}
3H^{2}= \rho+ \rho _{e}
\end{equation}
and
\begin{equation}\label{eq10}
2\dot{H}= -(\rho+ p)- (\rho _{e}+ p_{e})
\end{equation}
where $\rho _{e}$ and $p_{e}$ denote the effective energy density and the effective pressure
 respectively. These are given by
\begin{equation}\label{eq11}
\rho _{e}= \frac{u}{2}\dot{v} ^{2}- 3H\frac{\dot{v}}{v}
\end{equation}
and
\begin{eqnarray}
p_{e} &=& \frac{u}{2}\dot{v} ^{2} - H\frac{\dot{v}}{v} - \frac{\Box v}{v} \nonumber \\
&=& \frac{u}{2}\dot{v} ^{2} + 2H\frac{\dot{v}}{v} + \frac{\ddot{v}}{v} \label{eq12}
\end{eqnarray}

It may be noted that for $v=1$, we get back standard Friedmann
equations. In fractal cosmology, the fractal function can be time
like or even space like. At classical level, these two types of
fractal functions yield different physics. However, at quantum
level there is no analogous space or time like fractal functions
\cite{Calcagni2}.Therefore, here as in Ref\cite{Sheykhi4},  we
take time like fractal. To proceed further, we need to specify
fractal function $v$. In what follows, in order to remain
 general, we choose following two types of fractal functions \cite{Sheykhi4}.\\\\
{\bf Type I:}\\\\
First we assume a power law form of the fractal function $v$ as,
\begin{equation}\label{eq13}
v= v_{0}t^{-\beta _{f}}
\end{equation}

where $v_{0}$ is an arbitrary constant and $\beta _{f}$ is the fractal dimension. The parameters
 $\alpha _{f}$ and $\beta _{f}$ are related as $\beta _{f} = n(1-\alpha _{f})$. Note that for an
 ultraviolet nontrivial fixed point $\alpha _{f} = \frac{2}{n}$ while  $\alpha _{f} = \frac{4}{n}$
 for infrared fixed point \cite{Calcagni1}. So, in a four-dimensional space ($n= 4$), $\alpha _{f}$
 ranges as $0< \alpha _{f} \leq 1$. Subsequently the equations (\ref{eq11}) and (\ref{eq12}) take the forms
\begin{equation}\label{eq14}
\rho _{e}= \frac{u}{2}\beta _{f} ^{2}v^2 _{0}t^{-2(\beta _{f} + 1)}+ 3H\frac{\beta _{f}}{t}
\end{equation}
and
\begin{equation}\label{eq15}
p_{e}= \frac{u}{2}\beta _{f} ^{2}v^2 _{0}t^{-2(\beta _{f} + 1)}- 2H\frac{\beta _{f}}{t}+ \frac{\beta _{f}(\beta _{f} + 1)}{t^2}
\end{equation}
\\\\
{\bf Type II:}\\\\
Here we have considered an exponential form to the fractal
function $v$ given by
\begin{equation}\label{eq16}
v= v_{0}e^{-\beta _{f} t}
\end{equation}
Hence, for this type of fractal functins, equations (\ref{eq11})
and (\ref{eq12}) become
\begin{equation}\label{eq17}
\rho _{e}= \frac{u}{2}\beta _{f} ^{2}v^2 _{0}e^{-2\beta _{f} t}+ 3H\beta _{f}
\end{equation}
and
\begin{equation}\label{eq18}
p_{e}= \frac{u}{2}\beta _{f} ^{2}v^2 _{0}e^{-2\beta _{f} t}- 2H\beta _{f} + \beta _{f} ^2
\end{equation}

\section{Thermodynamical Analysis}

 In this section, we extend the idea of generalized Hawking temperature\,$(T^h)$ and modified entropy\,$(S_h)$
 to fractal universe at event horizon. In what follows, we study the followin three different generalized
 Bekenstein formulation at event horizon, namely
 \cite{Chakraborty3}:

({\it i}) $T^{h}= \frac{\alpha R_E}{2\pi R^2 _A}$ , $S_{h}= \frac{\pi R^2 _E}{G}$ .

({\it ii}) $T^h=\frac{R_E}{2\pi R_A^2}$ , $S_{h}= \beta\frac{\pi R^2 _E}{G}$ , $\beta = \frac{2}{R_E ^2} \int \frac{R_E^2 dR_A}{R_A}$ .

({\it iii}) $T^h=\frac{\alpha R_E}{2\pi R_A^2}$ , $S_h=\beta \frac{\pi R_E^2}{G}$ .\\\\
From the equations (\ref{eq11}) and (\ref{eq12}) we found
\begin{equation}\label{eq19}
\frac{\partial}{\partial t}(\rho _{e}+ p_{e})= \frac{\dot{v}}{v^2}(H\dot{v}- \ddot{v})-
 \frac{1}{v}(\dot{H}\dot{v}+ H\ddot{v})+ 2u \dot{v}\ddot{v}+ \frac{\dddot{v}}{v}
\end{equation}
using which in eq.\,(\ref{eq10}), one can get
\begin{equation}\label{eq20}
\frac{\partial}{\partial t}(\rho+ p)= 2f_{A}H^2+ 4v_{A}H\dot{H}- \frac{\dot{v}}{v^2}(H\dot{v}- \ddot{v})+
 \frac{1}{v}(\dot{H}\dot{v}+ H\ddot{v})- \left(2u \dot{v}\ddot{v}+ \frac{\dddot{v}}{v}\right)
\end{equation}
where $v_{A}= \dot{R} _A$, $f_{A}= \dot{v} _A$ .

\section*{Case-1}

Here we consider Bekenstein entropy  and the generalized Hawking temperature at the event
 horizon \cite{Chakraborty3} i.e.,
\begin{equation}\label{eq21}
S_{h}= \frac{\pi R^2 _E}{G}
\end{equation}
\begin{equation}\label{eq22}
T^{h}= T^{m}= \frac{\alpha R_E}{2\pi R^2 _A}
\end{equation}
where $\alpha= \frac{\dot{R} _{A}/R_A}{\dot{R} _{E}/R_E}$ is the reciprocal of the relative
 growth rate of the radius of the event horizon to that of the apparent horizon.

We now use the equation of continuity but in a modified form due to a fractal universe \cite{Sheykhi4} as
\begin{equation}\label{eq23}
\dot{\rho}+ \left(3H+ \frac{\dot{v}}{v}\right)(\rho+ p)= 0
\end{equation}

Clearly we can see, for $v=1$, equation\,(\ref{eq23}) reduces to the standard equation of continuity.
 Here the fractal is taken to be time-like only, so that the fractal function
 depends only on time {\it i.e.} $v= v(t)$. Therefore, considering the Gibb's relation (\ref{eq5}), we obtain
\begin{equation}\label{eq24}
dS_{f}= \frac{4\pi R_E ^2}{T^m}(\rho + p)\left(1+ \frac{\dot{v}R_{E}}{3v}\right)dt
\end{equation}
where we have considered the bounded fluid distribution with spherical volume $V_{h}= \frac{4}{3}\pi R^3 _E$.
 Using the equations (\ref{eq21}) and (\ref{eq24}), and involving eq.\,(\ref{eq23}), the time variation
 of the total entropy is given by
\begin{equation}\label{eq25}
\dot{S} _{T}= \frac{2\pi R_{E}\dot{R} _{E}}{G}- \frac{8\pi ^{2}v_{E}(\rho + p)}{v_{A}H^3}\left(1+ \frac{\dot{v}R_{E}}{3v}\right)
\end{equation}
and thus the second time derivative of the total entropy is given by
\begin{eqnarray}\label{eq26}
\ddot{S} _{T} &=& \frac{2\pi}{G}(R_{E}f_{E}+ v^{2} _{E}) - \frac{8\pi}{(v_{A}H^3)^2}\left[\left\{(v_{A}H^3)\left(f_{E}(\rho + p)+ v_{E}\frac{\partial}{\partial t}(\rho + p)\right)\right.\right. \nonumber \\
&-& \left.\left. v_{E}(\rho + p)(f_{A}H^{3}+ 3v_{A}H^{2}\dot{H})\right\}\left(1+ \frac{\dot{v}R_{E}}{3v}\right)+ \frac{v_{A}v_{E}H^{3}(\rho + p)}{3v^2}\left\{v(\ddot{v}R_{E}+ \dot{v}v_{E})- \dot{v} ^{2}R_E\right\}\right] \nonumber \\
\end{eqnarray}


where $v_{E}= \dot{R} _E$ and $f_{E}= \dot{v} _E$ .

\section*{Case-2}

In this case the horizon entropy is modified as \cite{Chakraborty3}
\begin{equation}\label{eq27}
S_{h}= \beta\frac{\pi R^2 _E}{G}
\end{equation}
where, $$\beta = \frac{2}{R_E ^2} \int \frac{R_E^2 dR_A}{R_A}$$
and Hawking temperature\,($=T^h$) is taken to be in the modified form
\cite{Chakraborty2}
\begin{equation}\label{eq28}
T^h=\frac{R_E}{2\pi R_A^2}.
\end{equation}
Here we can write $\beta$ as $$\beta=\frac{2}{R_E^2}\int \frac{R_E^2 dR_A}{R_A}=\frac{2}{R_E^2}\int \frac{R_E^2 v_A}{R_A}dt$$
 and from Gibb's relation (\ref{eq5}) we have
\begin{equation}\label{eq29}
\dot{S}_f=-\frac{8\pi^2 R_E (\rho+p)}{H^2}\left(1+ \frac{\dot{v}R_{E}}{3v}\right)
\end{equation}
Hence the first order time variation of the total entropy is
\begin{equation}\label{eq30}
\dot{S}_T=\frac{2\pi R_E^2 v_A}{GR_A}-\frac{8\pi^2 R_E (\rho+p)}{H^2}\left(1+ \frac{\dot{v}R_{E}}{3v}\right),
\end{equation}
and the second time derivative of the total entropy is
\begin{eqnarray}\label{eq31}
\ddot{S}_T &=& \frac{2\pi}{GR_A^2}\left[R_A R_E^2f_A+2R_AR_Ev_Av_E-R_E^2v_A^2\right]-\frac{8\pi^2}{H^4}\left[\left\{H^2\left(v_E (\rho+p)+R_E \frac{\partial(\rho+p)}{\partial t}\right)\right.\right. \nonumber \\
&-& \left.\left.2R_E(\rho+p)H\dot{H}\right\}\left(1+ \frac{\dot{v}R_{E}}{3v}\right)+\frac{H^2 R_E (\rho +p)}{3v^2}\left\{v(\ddot{v}R_E+\dot{v}v_E)-\do{v}^2R_E\right\}\right].
\end{eqnarray}

\section*{Case-3}

Finally, in this case  we take the horizon entropy as
\cite{Chakraborty3}
\begin{equation}\label{eq32}
S_h=\beta \frac{\pi R_E^2}{G}
\end{equation}
and the generalized Hawking temperature as \cite{Chakraborty3}
\begin{equation}\label{eq33}
T^h=\frac{\alpha R_E}{2\pi R_A^2}.
\end{equation}
Then the first time derivative of the total entropy is
\begin{equation}\label{eq34}
\dot{S}_{T}=\frac{2\pi R_E^2 v_A}{GR_A} - \frac{8\pi ^{2}v_{E}(\rho + p)}{v_{A}H^3}\left(1+ \frac{\dot{v}R_{E}}{3v}\right).
\end{equation}
The second time derivative of the total entropy is given by
\begin{eqnarray}\label{eq35}
\ddot{S} _{T} &=& \frac{2\pi}{GR_A^2}\left[R_A R_E^2f_A+2R_AR_Ev_Av_E-R_E^2v_A^2\right] - \frac{8\pi}{(v_{A}H^3)^2}\left[\left\{(v_{A}H^3)\left(f_{E}(\rho + p)+ v_{E}\frac{\partial}{\partial t}(\rho + p)\right)\right.\right. \nonumber \\
&-& \left.\left. v_{E}(\rho + p)(f_{A}H^{3}+ 3v_{A}H^{2}\dot{H})\right\}\left(1+ \frac{\dot{v}R_{E}}{3v}\right)+ \frac{v_{A}v_{E}H^{3}(\rho + p)}{3v^2}\left\{v(\ddot{v}R_{E}+ \dot{v}v_{E})- \dot{v} ^{2}\dot{R} _{E}\right\}\right] \nonumber \\
\end{eqnarray}

As in all the three cases the time variation of the total entropy are very complicated, so we
 cannot definitely conclude about their sign analytically. Hence we plot these time variation
 of entropies  $\dot{S} _{T}$ and $\ddot{S} _{T}$. For simplicity we consider the universe
 filled with perfect fluid having a constant equation of state {\it i.e.} $p= \omega \rho$.\\\\


\begin{figure}[h]
\begin{minipage}{0.4\textwidth}
\includegraphics[width=1\textwidth]{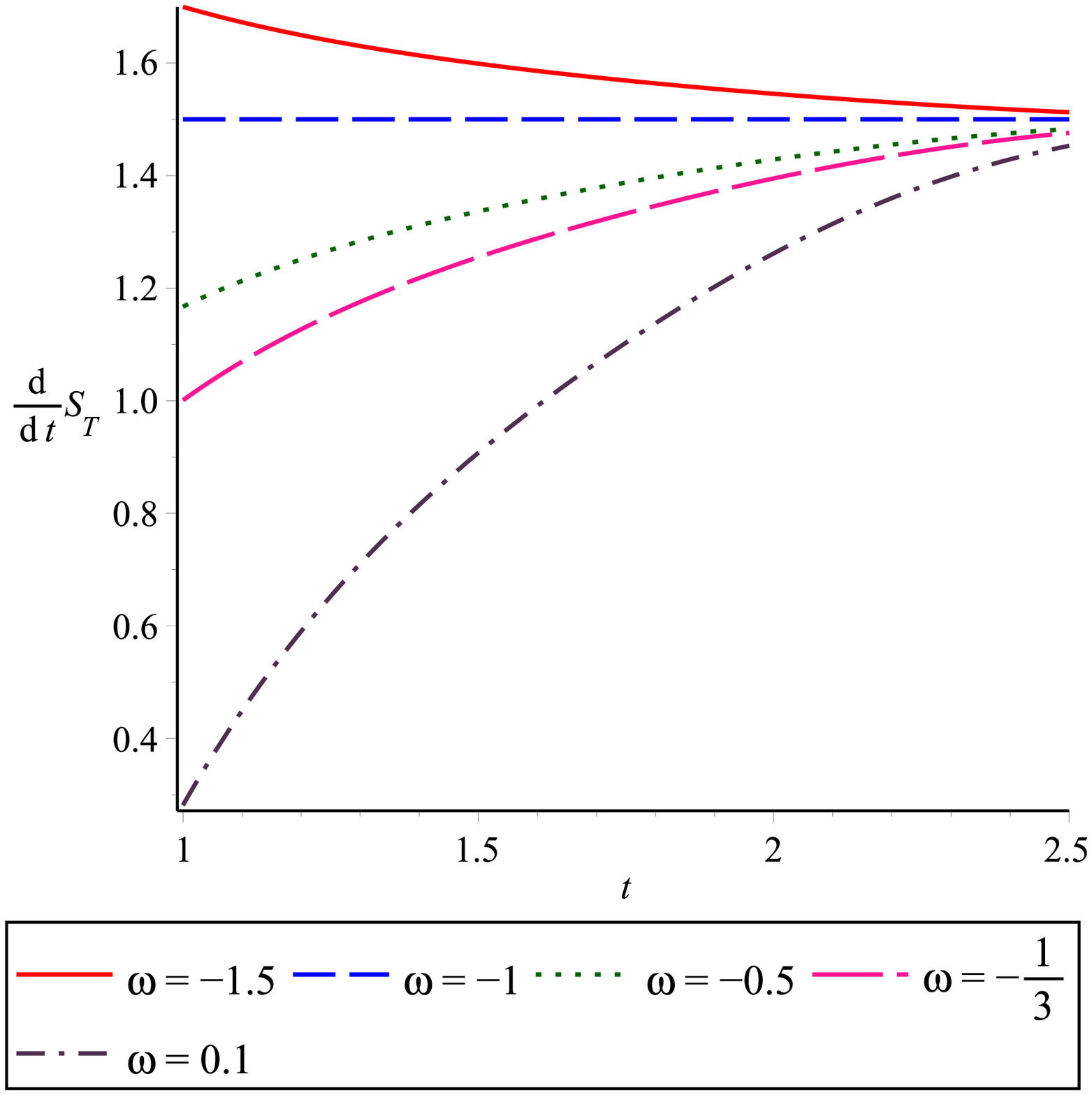}\\
Fig.1 : The time derivative of the total entropy is plotted against $t$ for Type-I in Case-1, considering $v_{0}= 5$ , $\beta _{f} = 3$ and $u = 10$.
\end{minipage}
\begin{minipage}{0.4\textwidth}
\includegraphics[width=1\textwidth]{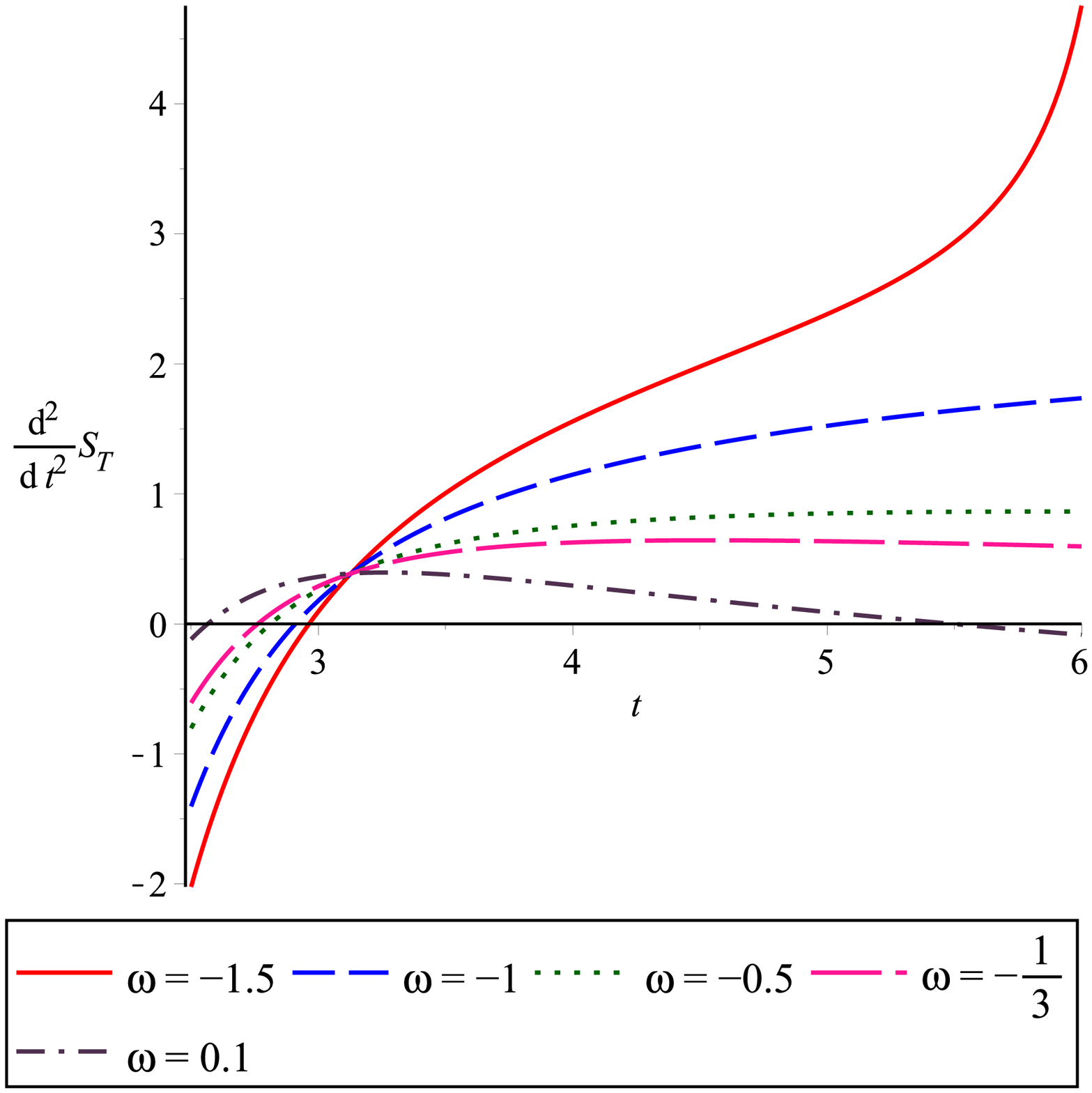}\\
Fig.2 : The second order time derivative of the total entropy is plotted against
 $t$ for Type-I in Case-1, considering $v_{0}= 5$ , $\beta _{f} = 3$ and $u = 10$.
\end{minipage}
\end{figure}

\begin{figure}[h]
\begin{minipage}{0.4\textwidth}
\includegraphics[width=1\textwidth]{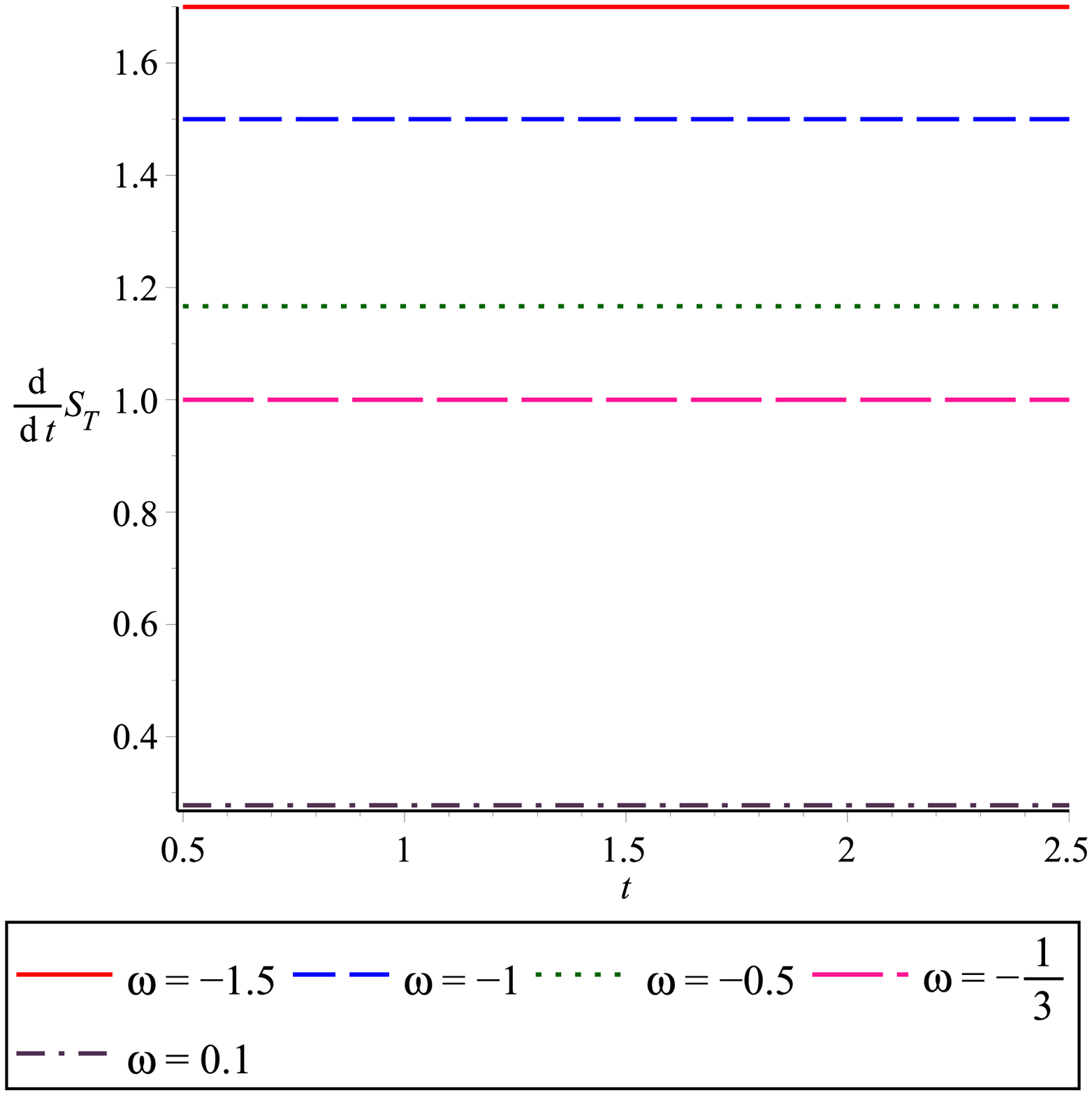}\\
Fig.3 : The time derivative of the total entropy is plotted against $t$ for Type-II in Case-1, considering $v_{0}= 5$ , $\beta _{f} = 3$ and $u = 10$.
\end{minipage}
\begin{minipage}{0.4\textwidth}
\includegraphics[width=1\textwidth]{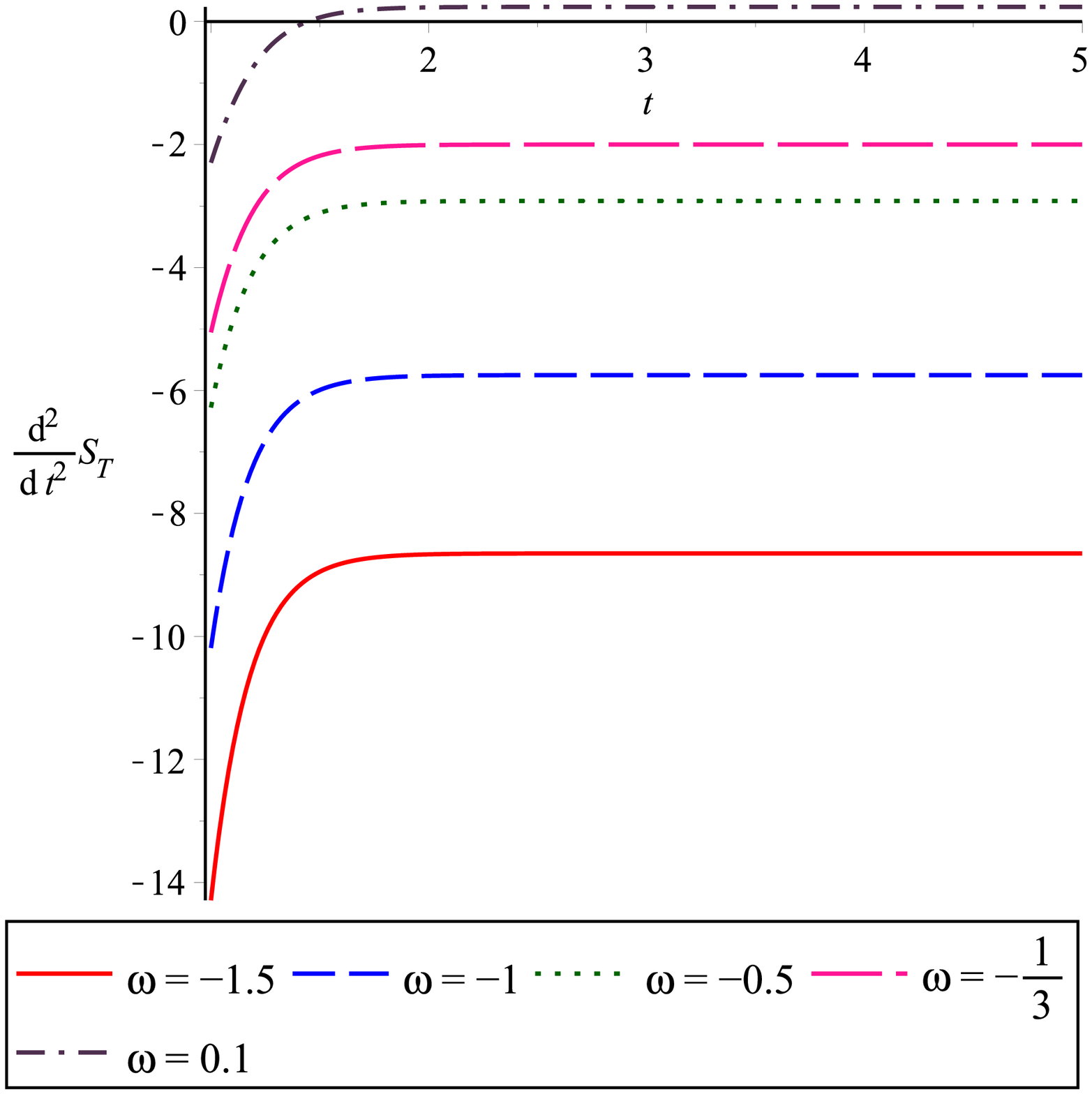}\\
Fig.4 : The second order time derivative of the total entropy is plotted against
 $t$ for Type-II in Case-1, considering $v_{0}= 5$ , $\beta _{f} = 3$ and $u = 10$.
\end{minipage}
\end{figure}


\begin{figure}[h]
\begin{minipage}{0.4\textwidth}
\includegraphics[width=1\textwidth]{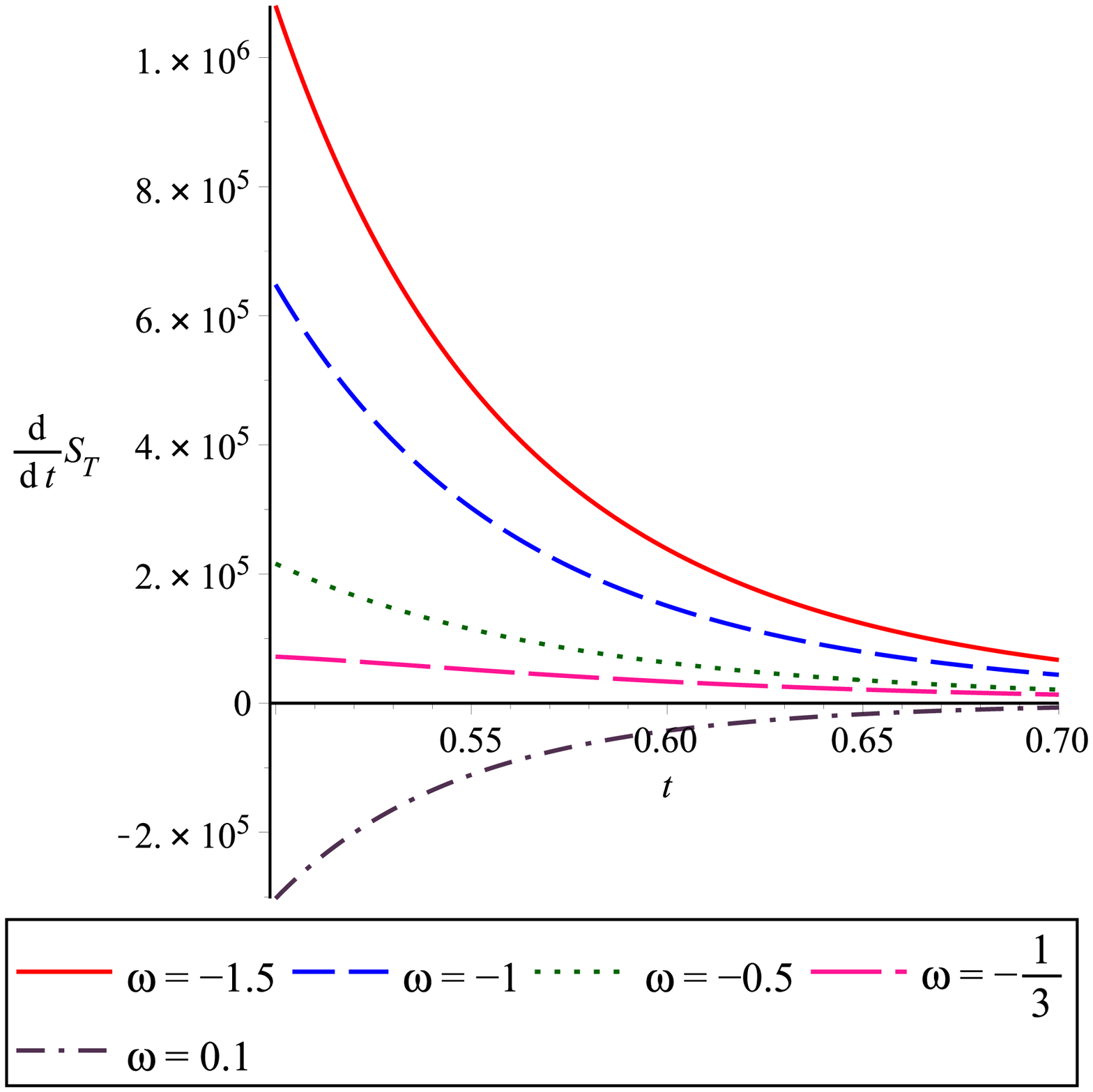}\\
Fig.5 : The time derivative of the total entropy is plotted against $t$ for Type-I in Case-2, considering $v_{0}= 5$ , $\beta _{f} = 3$ and $u = 10$.
\end{minipage}
\begin{minipage}{0.4\textwidth}
\includegraphics[width=1\textwidth]{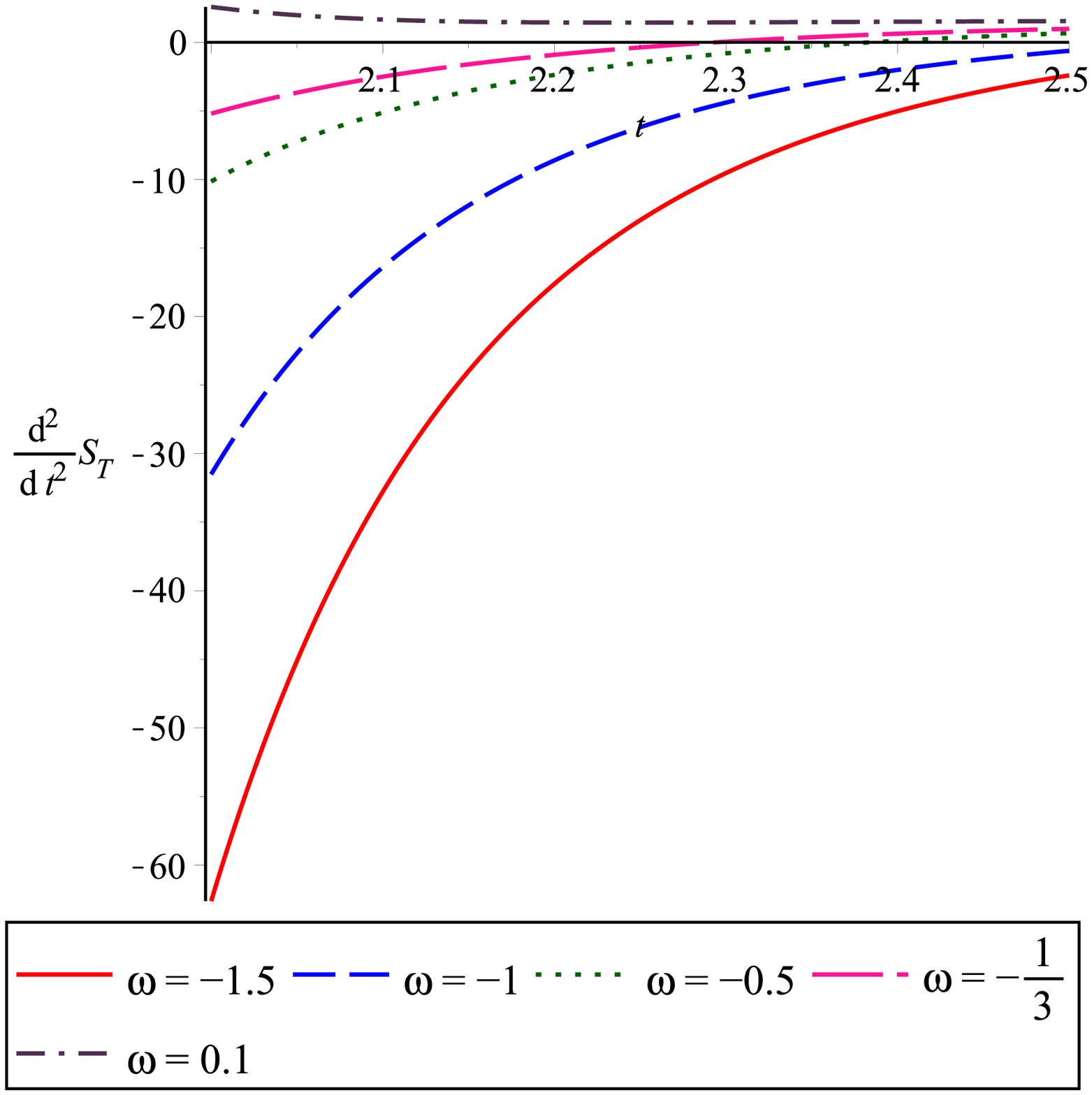}\\
Fig.6 : The second order time derivative of the total entropy is plotted against
 $t$ for Type-I in Case-2, considering $v_{0}= 5$ , $\beta _{f} = 3$ and $u = 10$.
\end{minipage}
\end{figure}

\begin{figure}[h]
\begin{minipage}{0.4\textwidth}
\includegraphics[width=1\textwidth]{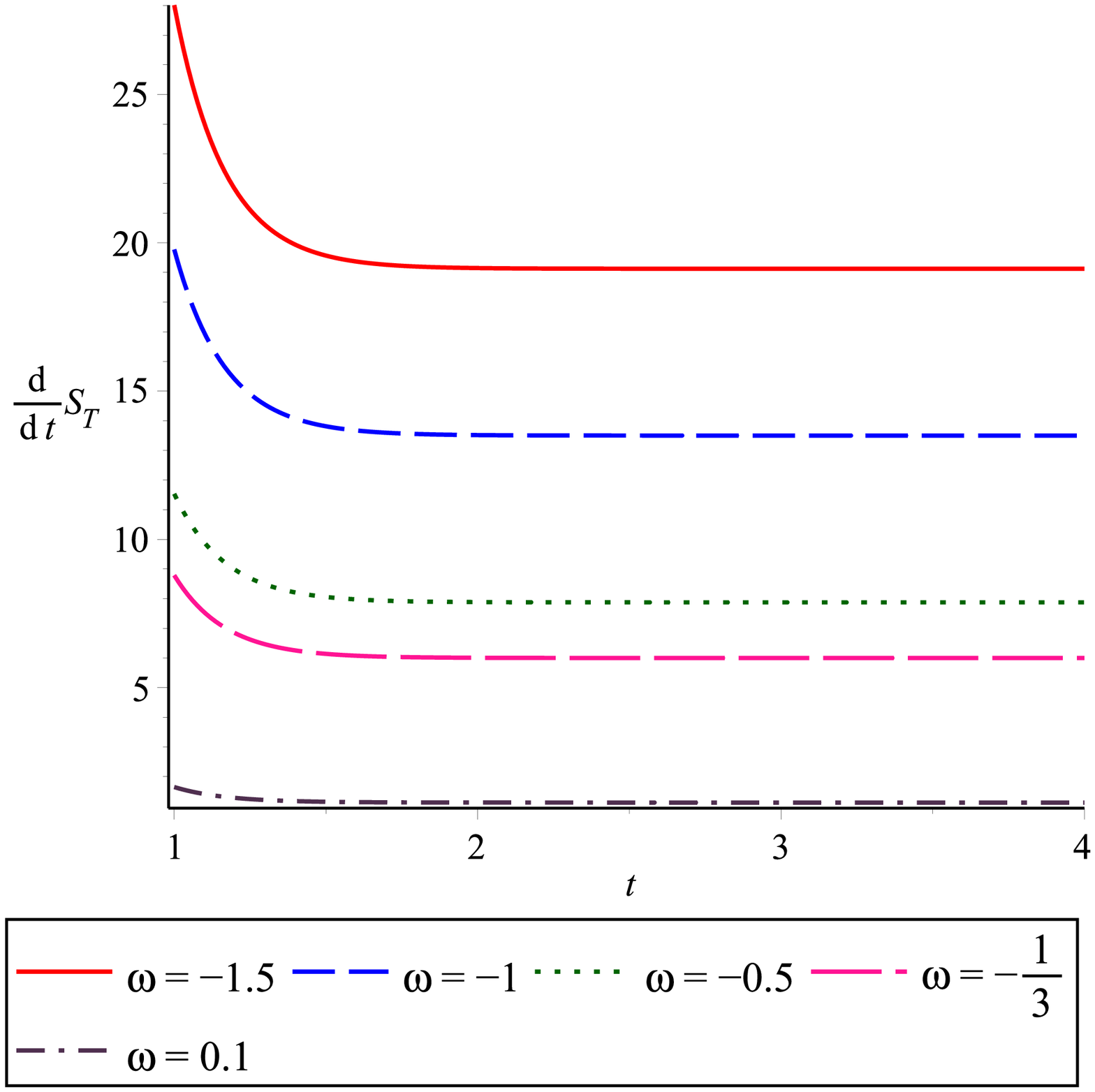}\\
Fig.7 : The time derivative of the total entropy is plotted against $t$ for Type-II in Case-2, considering $v_{0}= 5$ , $\beta _{f} = 3$ and $u = 10$.
\end{minipage}
\begin{minipage}{0.4\textwidth}
\includegraphics[width=1\textwidth]{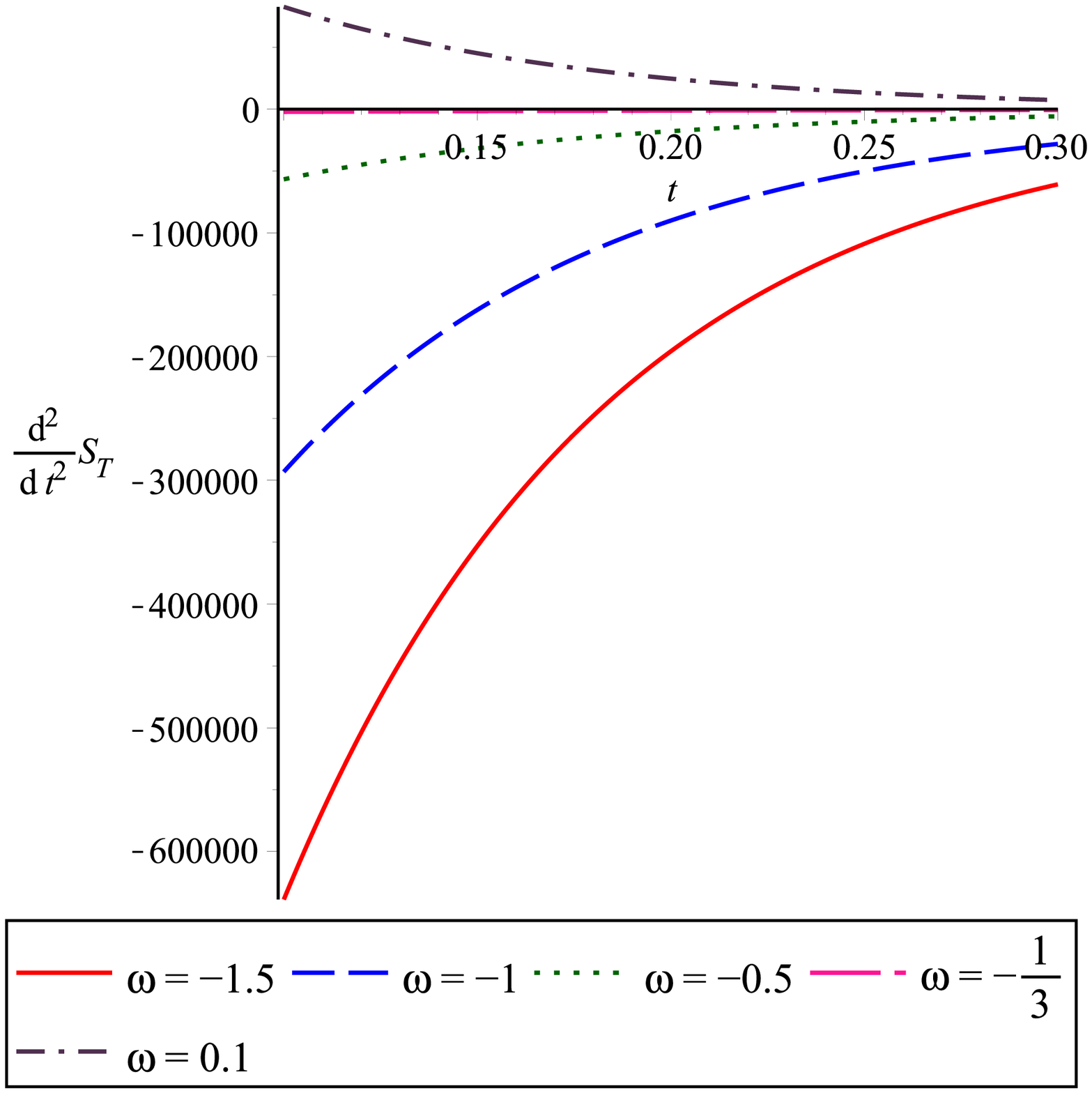}\\
Fig.8 : The second order time derivative of the total entropy is plotted against
 $t$ for Type-II in Case-2, considering $v_{0}= 5$ , $\beta _{f} = 3$ and $u = 10$.
\end{minipage}
\end{figure}




\begin{figure}[h]
\begin{minipage}{0.4\textwidth}
\includegraphics[width=1\textwidth]{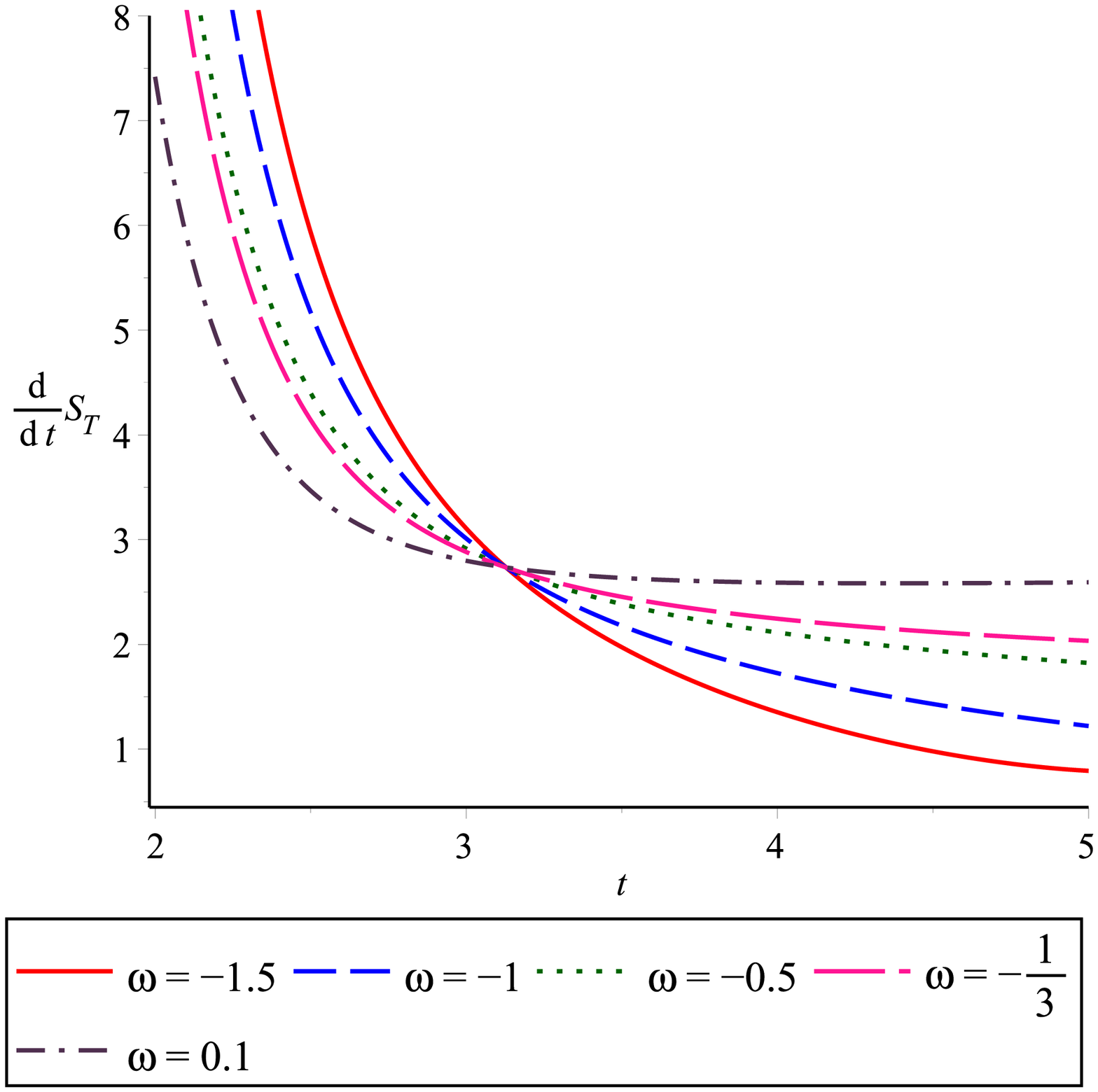}\\
Fig.9 : The time derivative of the total entropy is plotted against $t$ for Type-I in Case-3, considering $v_{0}= 5$ , $\beta _{f} = 3$ and $u = 10$.
\end{minipage}
\begin{minipage}{0.4\textwidth}
\includegraphics[width=1\textwidth]{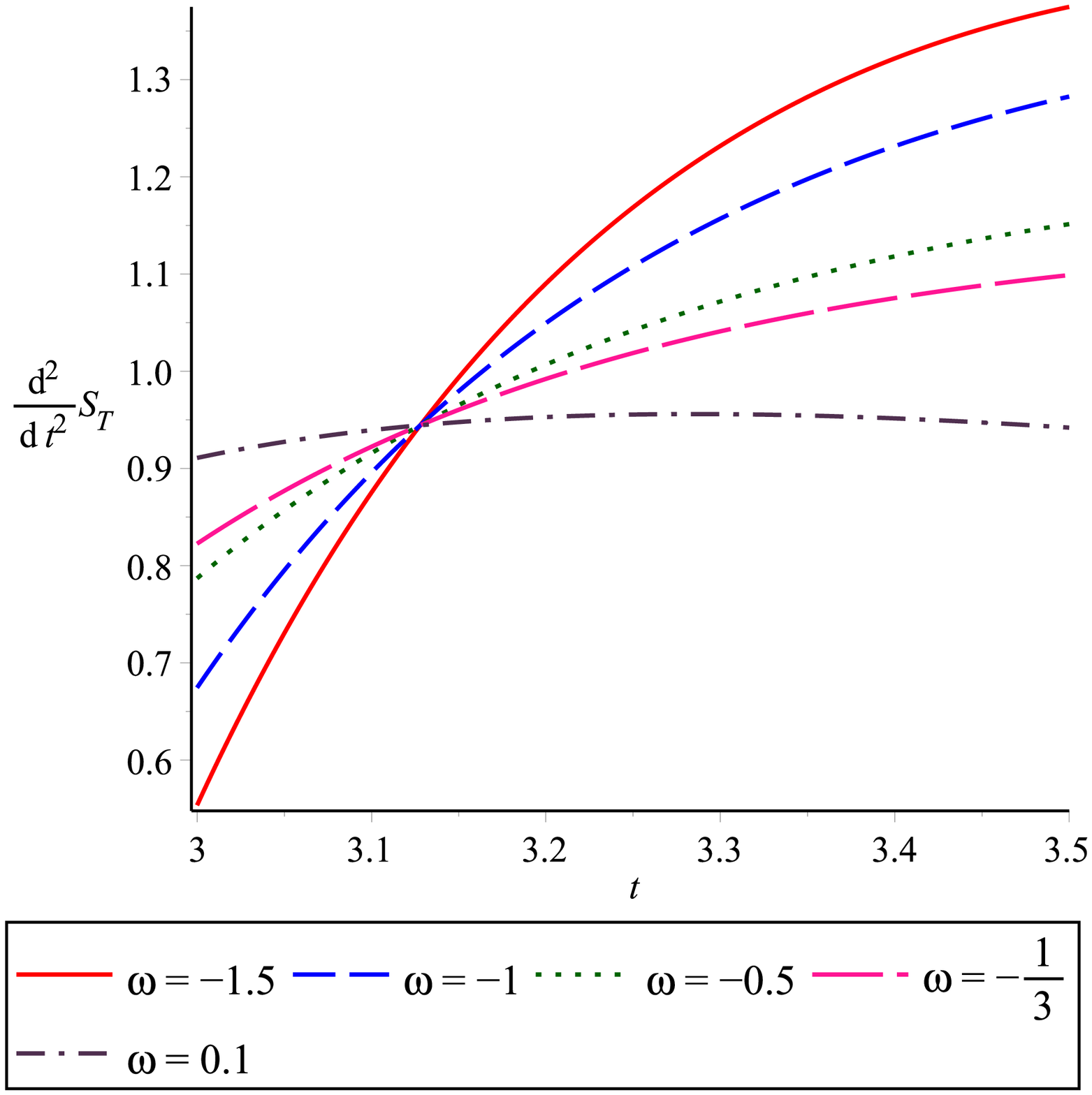}\\
Fig.10 : The second order time derivative of the total entropy is plotted against
 $t$ for Type-I in Case-3, considering $v_{0}= 5$ , $\beta _{f} = 3$ and $u = 10$.
\end{minipage}
\end{figure}

\begin{figure}[h]
\begin{minipage}{0.4\textwidth}
\includegraphics[width=1\textwidth]{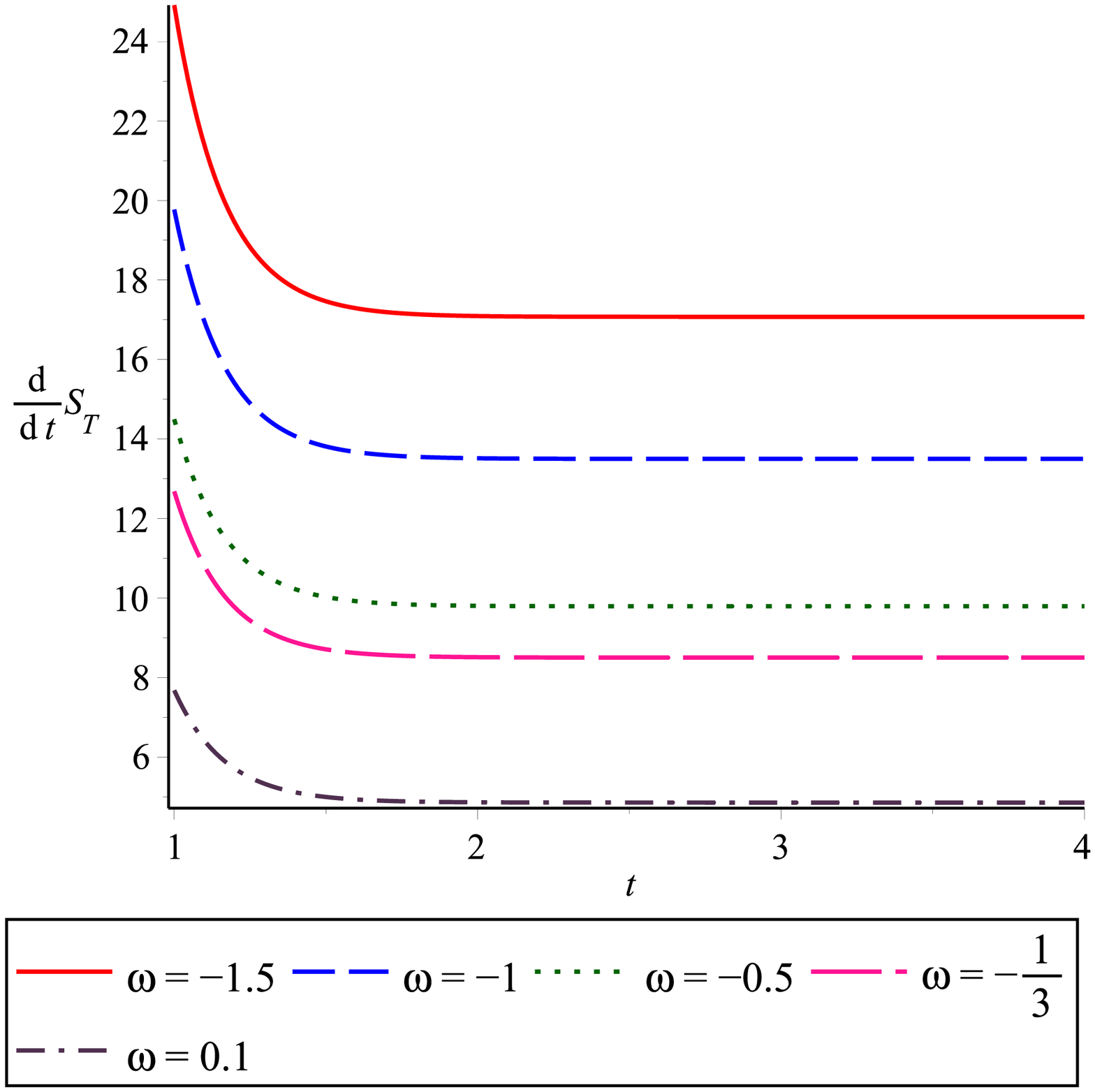}\\
Fig.11 : The time derivative of the total entropy is plotted against $t$ for Type-II in Case-3, considering $v_{0}= 5$ , $\beta _{f} = 3$ and $u = 10$.
\end{minipage}
\begin{minipage}{0.4\textwidth}
\includegraphics[width=1\textwidth]{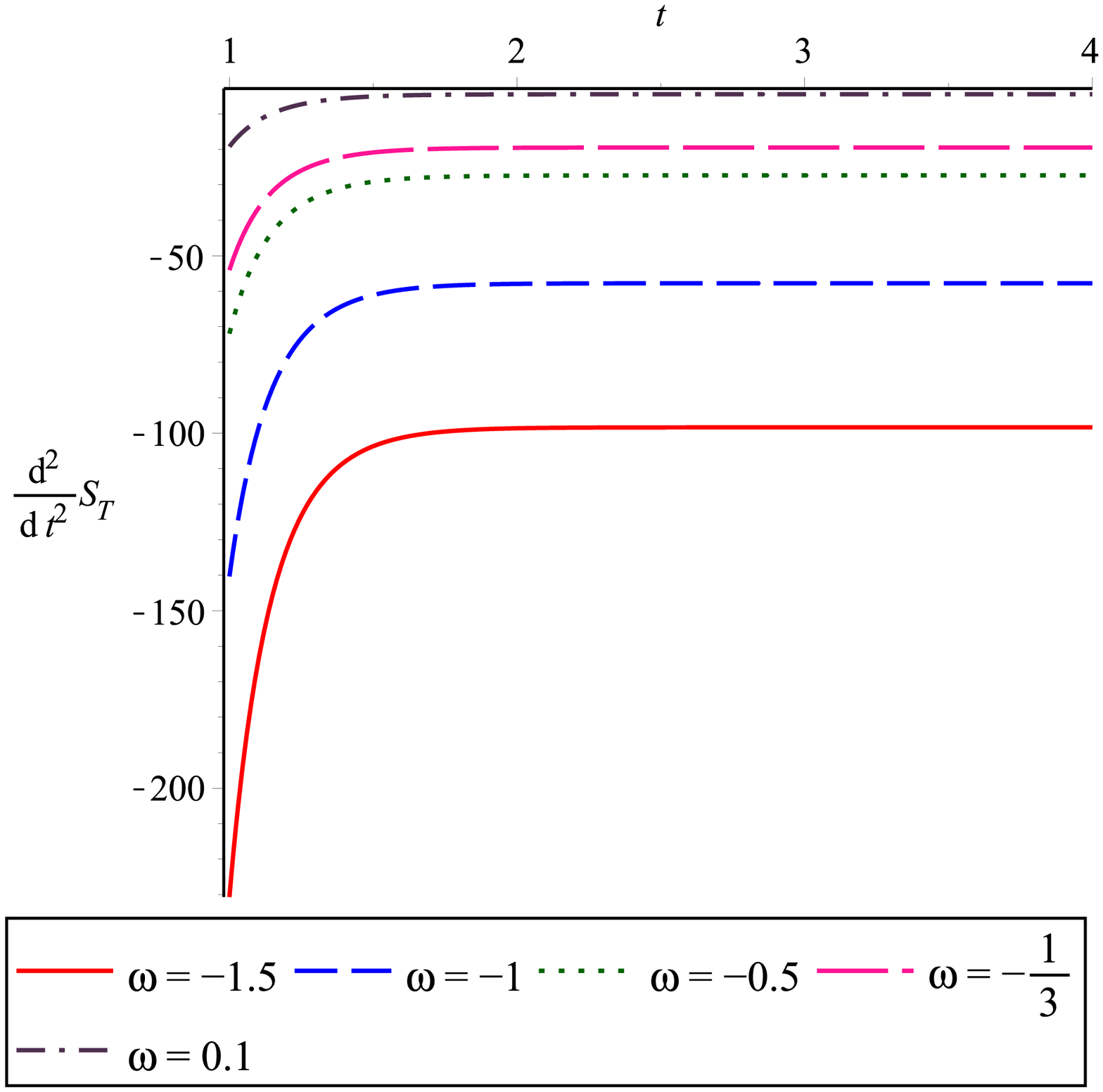}\\
Fig.12 : The second order time derivative of the total entropy is plotted against
 $t$ for Type-II in Case-3, considering $v_{0}= 5$ , $\beta _{f} = 3$ and $u = 10$.
\end{minipage}
\end{figure}

  From the above figures\,(1$-$12), we have the following observations:

  1) Both GSLT and TE hold for a wider range when the fractal function is an exponential function
  of time ({\it i.e.} Type-I) rather than power law form ({\it i.e.} Type-II).

  2) In most of the cases, GSLT and TE hold good when the values of equation of state parameter is
  taken to be negative, {\it i.e.} $\omega < 0$.

  3) From all the cases presented above, we have observed that third choice gives better result than
  the other two cases.\\


\section{Discussion}

In the present work we have considered two different types of fractal function $v$ and we have
 examined GSLT and TE  for both cases taking three different combinations of (modified) Hawking
 temperature and (modified) Bekenstein entropy. Due to complicated expressions of the time variation
 of the total entropy we cannot definitely conclude about validity of GSLT and TE. However, we have
 drawn some inferences only from graphical analysis considering $8\pi G= 1~,~~H= 1~~\textrm{and}
 ~~R_{E}=3$.

From the figures 1 and 2 we see that although GSLT is valid but TE is satisfied only for very
 early phase for the first choice of the entropy and temperature and with power law form of the
 fractal function and various choices of the equation of state parameter $\omega$\,. However,
 from figure 2, we also see that thermodynamical equilibrium will again be satisfied at late
 time for the choice $\omega = 0.1$\,. Figures 3 and 4 present the first and second order time
 derivative of the total entropy function for the first choice of the thermodynamical parameters
 with exponential form of the fractal function. Here both GSLT and TE are satisfied for all the
 choices of $\omega$\,(TE is marginally satisfied for $\omega = 0.1$). So one may conclude that
 exponential form of the fractal function is much favourable than polynomial form for
 thermodynamical study for the first choice of the thermodynamical parameters.

For the second choice of the entropy and temperature figures\,(5, 6) and figures\,(7, 8) show that
 the graphical variation of the first and second derivative of the total entropy function for the
 polynomial and exponential form of the fractal function. Here also the choice $\omega = 0.1$ is
 not thermodynamically viable as both GSLT and TE are not obeyed\,(at least marginally).

Lastly, for the third choice of the thermodynamical parameters the graphical representations\,
(in figures 9--12) show that there will no longer be any TE for the power law form of the fractal
 function while both GSLT and TE are satisfied\,(marginally for $\omega = 0.1$) for exponential
 form of the fractal function.

Therefore, from the above graphical analysis one may conclude that second choice of modified
 Hawking temperature and modified Bekenstein entropy is much favourable for the study of universal
 thermodynamics in fractal universe and exponential form of fractal function is more realistic
 from the point of view of GSLT and TE. Also negative value of $\omega$ is suitable for these
 thermodynamical studies.

Finally, we may conclude that modified Bekenstein entropy and modified Hawking temperature
 can be considered as realistic thermodynamical parameters on the event horizon of the fractal
 universe.\\\\
\newline
{\bf Acknowledgement:}\\

The authors are thankful to IUCAA, Pune, India for research facilities at Library. Also
 SC acknowledges the UGC-DRS Programme in the Department of Mathematics, Jadavpur University.\\\\


\end{document}